# Architectural Design Activities for JAS

Lena Khaled
Software engineering department
Zarqa private university
Amman,Jordon
lenaumleen@yahoo.com

*Abstract—the critical part for building any software system is its architecture. Architectural design is a design at a higher level of abstraction. A good architecture ensures that software will satisfy its requirement. This paper defines the most important activities of architectural design that used through building any software; also it applies these activities on one type of Electronic Commerce(EC) applications that is Job Agency System(JAS) to show how these activities can work through these types of applications.*

## I. INTRODUCTION

Architectural design has been an important subfield of software engineering. According to IEEE "Software architecture (also called architectural design) is the fundamental organization of system embodies in its components, their relationships to each other and to other environments, and the principles guiding design and evolution". So, architecture design is a design at a higher level of abstraction and it focuses on externally visible properties of software elements [8, 7]. The aim of this paper is to define the main activities of building the architectural design to any software system; these activities of architecture are applied on Job Agency System (JAS) as one type of electronic commerce applications which is the case study of this paper.

The paper is organized as the following: section 2 shows an overview of related works on architectural design with electronic services, Section 3 gives definitions to architectural activities, Section 4 defines the case study; Section 5 shows how architectural design activities can be applied on JAS and the conclusion of this paper is done through section 6.

## II. RELATED WORKS

Many different researches work on architectural design with electronic commerce in many different ways: Perry and Wolf [6] work on the foundations for the study of software architecture in general. They present a model with three components: element, form and rationale. They define elements as either data or processing, forms are either the properties of elements or constraints on them and finally they define rationale as the basis of architecture in terms of constraints. Widhain and his group [4] work on building pattern for electronic commerce system they identify and derive patterns on architectural level that are specified on to the domain of electronic system. This work defines architectural design activities on one type of electronic commerce application that Job Agency System (JAS) as described in the following section.

## III. Architectural design activities

Architecture is concerned with the selection of software elements, their interaction and the constraints on these elements to provide framework that serves as a basis for a design and satisfy the requirements of building [5]. Architectural design typically plays the key role between requirement and implementation; it provides an abstract description of a system as shown in the following figure [3].

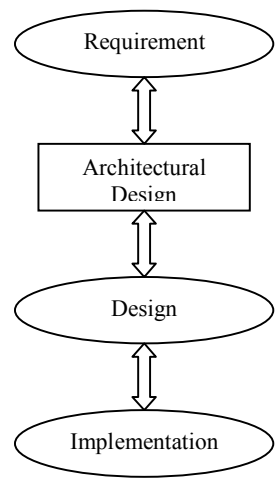

Figure 1. Architectural design as a Bridge

To build architecture to any software system, three activities must be described. These activities are shown in figure2 and described in the following subsections.

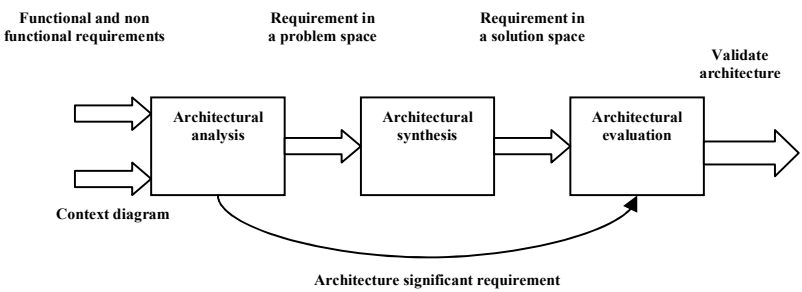

Figure2. Architectural design activities

### A. Architectural analysis

This activity analyzes the architecture to define the main problem. It also models the user's requirements for what the system must do. This step has many advantages, for example: easy maintain system, rapid development and ease the transition to the design phase.

As shown in figure 2, architecture analysis needs the following important steps:

#### 1. Project scope and context diagram:

The project scope is a statement of what the project produces, it tells in a brief statement what the eventual system will do. While the purpose of context diagram shows how the system relates to the world beyond itself, it models the flow of data to and from the external entities which the system interacts with [2].

#### 2. Functional and non functional requirements.

Functional requirements are statements of services that should be offered by the system to the users. It shows how the system reacts to specific inputs and the behavior of the system in particular situations. Non-functional requirements are defined as constraints on the functions offered by the system. They include security constraints, timing constraints, constraints on the development process and standards such as performance and scalability. Non-functional requirements often apply the system as a whole; they don't apply specific features or services [9].

### B. Architectural synthesis

This activity is the core of architectural design activities; it processes the architecture requirements to a set of Architectural Significant Requirements(ASR), so it moves the requirements from problem space to requirements in a solution space. Architectural Significant Requirements (ASR) can be defined as a set of requirements upon a software system which influences its architecture [4].

### C. Architectural evaluation

This last activity ensures that the proposed architecture is the right one. The proposed architecture is measured through significant requirements. Multiple iterations are expected through this activity to ensure the validity of building architecture.

Validation is a task to ensure that software does what its users want. Architectural validation consists of those proposed architectural solutions which are consistent with ASRs. Only one architecture chooses which is the most appropriate to the significant requirements [4, 1].

## IV. The case study (*JA*S)

Job Agency System (JAS) is one type of Businesses-to-Consumer (B2C) commerce which is the type of commerce that includes retail transactions of services from business to individuals. JAS is effective for technology oriented because this type of systems use internet regularly. Traditional JAS is different from electronic JAS in the ability of employees to compare jobs; traditional JAS is very limited while electronic JAS is very easy and fast. The following section describes how the previous activities can be applied to build architecture to such type of systems.

## V. Architectural design activities for JAS

### A. Architectural analysis

The main problem of the system is defined through this step and then models the user's external entities that interact with JAS. The last activity in this step is to define the main tasks for each entity through a use case diagram as shown in the following.

#### 1. The Project scope of JAS

Project scope is a group of statements which generally describe what JAS will do, what functions will be part of it and which users will be served as described in the following figure

> JAS is a system which connects individuals who are looking for jobs the employer (recruiter) who is looking for employee with specific skills. This type of systems serves especially people who are looking for jobs.

Figure3. Project scope for JAS

*2. The context diagram of JAS*

This section defines all the external entities outside the system, which will either feed data to it or need to get data out of it as in the following figure.

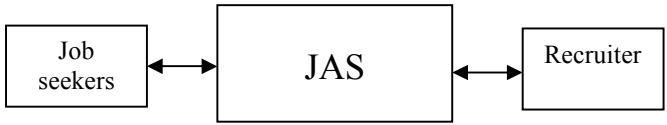

Figure4. Context diagram for JAS

In figure 4, the system seems to be as a black box and the most important entities which interact with it are individuals (job seekers) and employers (recruiter). The job seeker feeds data to the system which is his C.V., while the recruiter feeds information which is the properties of specific skills that are requested by market. Both, job seekers and recruiters, are matching around the clock through a supplier's electronic store as we see in synthesis part that is the second step of architecture activity.

*3. Use case diagram of JAS*

In Figure 4, the recruiter and the job seeker are the main users to the system. In this section the main tasks of each one are described.

*A  Recuriter*

A recruiter is a user that interacts with JAS; he is the person who works for an organization to recruit candidates for different empty posts in it.

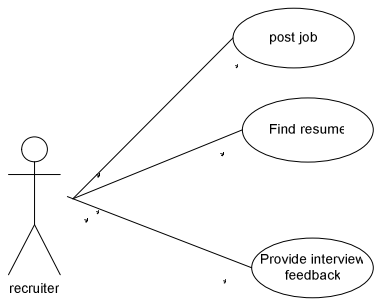

Fig 3. The use cases of recruiter

The descriptions of each operation are described in the following table:

Table 1. Descriptions of the recruiter's operations

| Operation | Description |
|---|---|
| Post job | Recruiter can put a new job in the web site. |
| Find resume | Recruiter can search database to find specific employee by searing criteria such as an employee type. |
| Provide interview feedback | Recruiter can check the online interviews and provide the answers of these interviews. |

*B. job seeker*

The job seeker is the second user of JAS that takes the initiative and places his resumes on his homepage or on other websites. The main tasks of a job seeker are shown in the following:

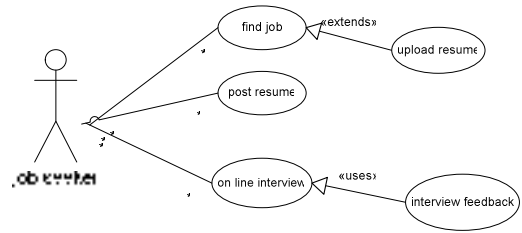

Fig 4. The use cases of a job seeker

The descriptions of each operation are described in the following table:

Table 2. Descriptions of job seeker operations

| operation | description |
|---|---|
| **find job** | Job seekers search data base by providing their ser search criteria. |
| **post resume** | Job seeker can post his resume for specified job. |
| **Upload resume** | A job seeker can upload his resume to the system for further use. |
| **Online interview** | A job seeker can give online interview for sharing the process of a specific job. |
| **Take interview feedback** | A job seeker that share on line interview can take a feed back about the result of an interview. |

*4. Non-functional requirement for JAS*

Non functional requirements represent the criteria that can be used to control the operation of any system. Sometimes non functional requirements are called the qualities of the system that are defined from the point of view of the user's goal or need. In JAS, the most important quality is scalability, which is defined under an evolution quality that is embodied in the static structure of the software system. Scalability is important to some e-commerce applications that need to make the most cost-effective use of their computing resources, and grow their computing environment along with their business.

Horizontal scaling is characterized by rapid application change and simplified change management. With multiple servers in horizontal tier, changes can be deployed incrementally enabling applications or services to be replicated quickly to multiple servers in a controlled manner. Vertical scaling services that are scaled within the system resources can be incrementally added to the server over time to increase scalability; it is characterized by slowly changing applications. Diagonal scaling is a combination between vertical and horizontal scaling providing increased flexibility.

*B. Architectural synthesis for JAS*

As described in previously, the architectural synthesis is the core of the architectural design. The architectural synthesis for JAS can be described in figure 5.

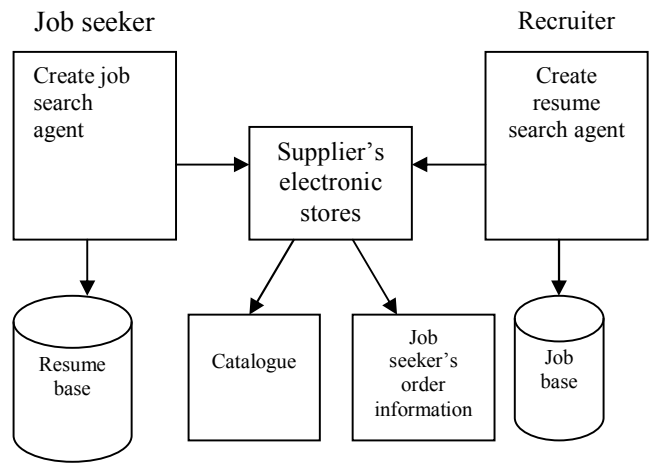

Fig 5. The architectural synthesis for JAS

figure 5 shows that the job seeker and recuriter are matching around the clock, and because of a larage number of available jobs and resumes online, the importance of intelligent agent (both to job seeker and recruiter) appears. Intelligent agents are autonomous software entities that can navigate through heterogeneous computing environments and can either work alone or with other agents to achieve the goal [11]. Any customer in electronic commerce interacts with various agents in the process of making decisions and sometimes there are some interactions among the software agents to complete the decisions. In this paper the type of agent that connect through is called a consultation agent that connects the job seeker and recruiter to the knowledge base as represented in figure 6.

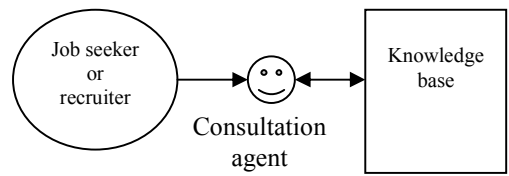

Fig 6. Consultation agent and it interactions

From figure 6 we see the important role of an intelligent agent for recruiter and job seeker, the job seeker can create many different profiles based on different job categories: geographic, regions and key words. The job seeker receives a daily e-mail that contains job opportunities from a dozen top job sites around the internet. This saves the job seeker a tremendous amount of time while the intelligent agent can help the recruiter finding resumes that match specific job descriptions.

The concept of a catalogue appears in figure 5. A catalogue is an important component of business process, and it might consist of a set of web pages that describe a specific job and its organization with embedded pictures and drawings. The web pages might be created as static pages using HTML from the database of the job and the descriptive information.

*C.* Architectural evaluation for JAS

The prototype architecture for JAS consists of three layers which is called sometimes tiers, this N tier architecture was designed to enable the distribution of tasks as possible. A tier (layer) is functionally a specific hardware and software component that performs a specific task. The communication between tiers is done through protocols. This N tier architecture enables the system components to be easily scaled. The prototype architecture for JAS that based on N tier is shown in figure 7.

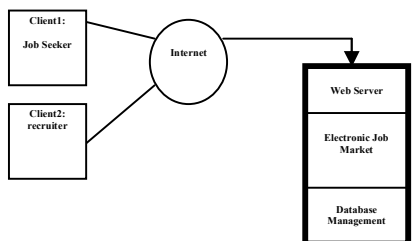

Fig 7. The architectural evaluation for JAS

Two major clients are communicated to the system through the web browser that are the job seeker and the recruiter. These are the layers of the system:

- Web server which contain a catalogue, a catalogue is an important components of business process, a web server may contain a catalogue server for a structured content for example a product catalogue or a job market catalogue that presents in some web directory.
- Application layer contains all processing that electronic market needs such settlement. In the settlement phase, monetary transactions occur between the client and the system.
- Data layer can be either relational database such as Access or object oriented database.

VI. CONCLUSION

This paper combines between software engineering through software architecture and e-commerce through JAS. s paper applied the activities of architectural design on JAS. It shows the external entities that interact with JAS and the main task of each one through analysis which is the first step of activities, then defines the basic structure of JAS through synthesis which is the second step of the activity and finally build the prototype of JAS through evaluation which is the third step of the activity.